# Submission for ACM Papers

CityScope: Enhanced Localization and Synchronizing AR for Dynamic Urban Weather Visualization

Bridging the Tangible and Virtual with AR on a Physical Representation

Tzu Hsin Hsieh

Researcher, Department of Architecture, Department of Computer Science, National Cheng Kung University, celinehsieh68@gmail.com

CityScope uses augmented reality (AR) to change our interaction with weather data. The main goal is to develop real-time 3D weather visualizations, with Taiwan as the model. It displays live weather data from the Central Weather Bureau (CWB), projected onto a physical representation of Taiwan's landscape.

A pivotal advancement in our project is the integration of AprilTag with plane detection technology. This innovative combination significantly enhances the precision of the virtual visualizations within the physical world. By accurately aligning AR elements with real-world environments, CityScope achieves a seamless and realistic amalgamation of weather data and the physical terrain of Taiwan.

This breakthrough in AR technology not only enhances the accuracy of weather visualizations but also enriches user experience, offering an immersive and interactive way to understand and engage with meteorological information. CityScope stands as a testament to the potential of AR in transforming data visualization and public engagement in meteorology.

## 1 INTRODUCTION

In the modern era, where data is not only abundant but also pivotal in decision-making, the challenge lies in effectively visualizing and interpreting this data. For instance, weather data, with its inherent complexity and variability, requires innovative visualization techniques to be fully comprehensible and actionable [1]. Traditional methods often fail to convey the dynamic nature of meteorological data, making it difficult for users, including professionals and the public, to grasp and utilize this information effectively [2].

The concept of localization in data visualization, particularly in Augmented Reality (AR), has become increasingly significant. Accurate localization ensures that digital information is not just overlaid but is contextually and spatially aligned with the physical world, enhancing comprehension and usability [3]. This is crucial in applications where decision-making relies on the precise geographical positioning of data, such as urban planning, environmental monitoring, and disaster management [4].

However, achieving precise localization in AR remains a challenge. The discrepancies in aligning virtual data with physical locations can lead to misinterpretations, especially in complex and dynamic fields like meteorology or urban planning [5]. These challenges highlight the need for advanced AR technologies and methodologies that can improve localization accuracy, thereby enhancing the reliability of data visualization in real-world applications [6].

Our application aims to address these challenges by leveraging the latest advancements in AR for enhancing the precision of data localization. By focusing on the accurate alignment of complex data sets, such as meteorological information, with their real-world counterparts, we strive to bridge the gap between data complexity and user comprehension. This approach is not only about visualizing data but ensuring its contextual relevance and spatial accuracy, thereby contributing to more informed decision-making and a deeper understanding of critical data sets [7].

In this paper, our goal is to develop an interactive interface that allows the public to understand information more intuitively. We have designed an enhanced localization system that maps visualized data onto a physical representation of Taiwan. Our objective is to synchronize the tangible and virtual realms by displaying real-time weather visualizations through an AR app. Additionally, the precise localization provided by our system offers urban decision-makers a more accurate understanding of city-specific information. This approach not only enhances public engagement with complex data but also aids in the effective planning and management of urban environments. By integrating advanced AR technologies with accurate data mapping, we create a tool that is both informative and accessible, bridging the gap between complex data sets and practical, real-world applications.

## 2 RELATED WORK

### 2.1 Augmented Reality in Data Visualization

The application of Augmented Reality (AR) in data visualization presents significant opportunities for enhancing civic engagement and urban decision-making processes. By integrating AR technology into mobile applications, data visualization becomes more interactive and immersive, allowing for a more engaging and informative experience for users. This is particularly relevant in urban settings where complex civic data needs to be presented in an accessible and understandable manner to encourage public participation. For example, a low-cost AR mobile app for data visualization empowers citizens to better understand urban data and engage in decision-making [8].

Furthermore, the evolving nature of AR in data visualization is evident in its application in various domains, including urban planning and policy-making. For instance, AR's capability to superimpose detailed 3D models onto real-world environments enhances the visualization of urban projects and future developments [9]. This approach not only provides planners and decision-makers with a more tangible understanding of potential outcomes but also allows for better communication and collaboration with the public. In turn, this fosters a more transparent and inclusive urban planning process, where citizens are better informed and can contribute more effectively to shaping their cities.

### 2.2 Advancements of AR Localization

The advancements in localization for Augmented Reality (AR) allow users to smoothly integrate digital enhancements within our physical reality. As depicted in Figure 1, there are two primary types of AR localization methods that facilitate this integration: Marker-Based AR and Markerless AR. Marker-based AR utilizes visual cues, such as QR codes, which when detected by an AR device, trigger the overlay of digital content onto these markers in the real world. This method, established by foundational work such as that by Wagner and Schmalstieg [11], allows for precise and stable augmentation. Meanwhile, projection-based AR, highlighted by Bimber and Raskar [12], uses light to transform physical surfaces into interactive displays, such as projecting a virtual keyboard onto a table, enabling new forms of user interaction with their immediate environment.

On the other hand, markerless AR technologies have opened up a realm of possibilities where physical markers are no longer necessary. Location-based AR, often relying on GPS, allows for the contextual placement of digital content, as first discussed in the seminal work by Azuma et al [13], and is particularly effective for navigation and providing information overlays in outdoor settings, such as platforms like ARKit and ARCore, as discussed by Linowes et al [14]. Contour-based AR, advanced by the likes of Newcombe et al. [15] with their KinectFusion project, enables the recognition of environmental shapes and outlines, allowing virtual objects to be placed in more accurate alignment with the real world, enhancing the realism and immersion of the AR experience.

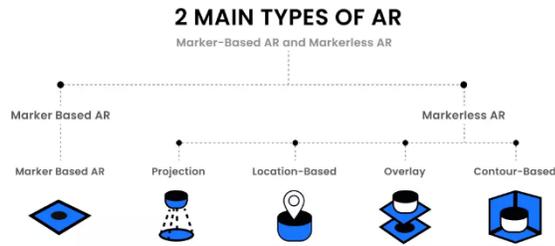

Figure 1: Types of AR localization methods [10]

### 2.3 Bridging Data Complexity and User Comprehension

The synergy between Data Visualization and AR Localization is central to overcoming the challenges posed by complex data sets, thereby significantly enhancing user comprehension. In this study, we have developed a system that combines the advantages of Marker-Based AR for precise positioning with the flexibility of Markerless AR, which maintains the placement of virtual objects even when the camera moves. This integration allows for the effective visualization of real-time meteorological data. These technologies ensure that data is not only visually intuitive but also geospatially relevant, empowering users to easily grasp sophisticated information. The ability to place abstract data into the physical realm where actions occur bridges the gap between knowledge and application, allowing for immediate and informed decision-making and enhancing public participation.

## 3 METHODS

### 3.1 System Design

The app harnesses open weather data from the Central Weather Bureau (CWB) of Taiwan, integrating crucial environmental metrics such as the UV index, rainfall, temperature, and PM2.5 levels provided by the Environmental Protection Agency (EPA).

As shown in Figure 2, at the heart of our application are two main visual components that work in tandem to provide both an immersive experience and detailed information:

1. Upper Sphere (Main Visual Element): This component is the centerpiece of our application.

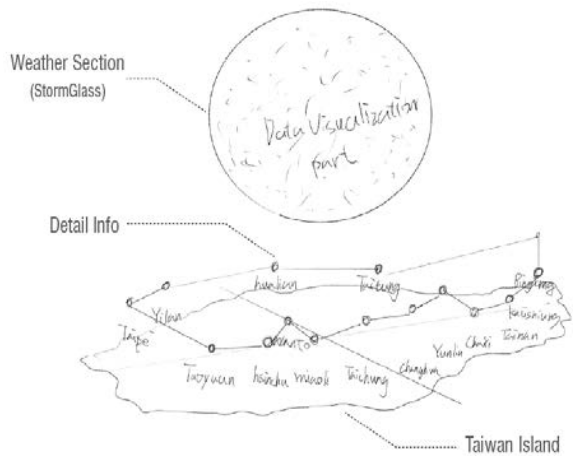

Figure 2: Concept of the System Design

Designed to resemble a globe weather, it offers an intuitive and engaging way to visualize weather conditions. Inside this sphere, users can see particle changes that vividly represent different weather phenomena, from the swirling of a typhoon to the subtle variations indicating changes in air quality. This dynamic representation goes beyond traditional weather maps, offering a more relatable and engaging way to understand complex meteorological data.

2. Lower Pin Object (Informational Aid): Complementing the upper sphere, the lower object serves as an informational aid. It displays critical data metrics in an easily digestible format, alongside the names of cities or counties. This feature is particularly useful for users seeking specific information about their local area or for those who wish to compare conditions across different regions.

The combination of these elements within an AR framework represents a significant advancement in data visualization and public engagement in meteorology. By localizing and synchronizing real-time data within an interactive AR environment, our application, CityScope, not only enhances the accuracy of weather visualizations but also enriches the user experience. It offers an immersive and interactive way to understand and engage with meteorological information, setting a new standard for how weather data can be presented and interacted with in the digital age.

In the next section, we will explore the technical foundations of our app, focusing on our approach for precise localization and the approach to data visualization. We'll discuss how users AprilTag enhances the integration of virtual and physical elements and how our visualization techniques make complex data more accessible and engaging for, highlighting the synergy between these two key aspects in augmenting the user experience.

### 3.2 Development Environment and Tools

In this study, we leveraged the Unity game development engine, version 2021.3.12f1, for the development of our prototype. This choice enabled us to craft a real-time weather observation application that is compatible with iOS systems. The application utilizes AR technology to enhance the user experience, featuring a virtual sphere superimposed on a physical model of Taiwan for an interactive display of weather data.

In addressing the challenge of precise AR localization, we integrated the AprilTag tracker. Developed by the APRIL Robotics Laboratory at the University of Michigan, AprilTag is a marker-based tracking system. We utilized the jp.keijiro.apriltag Unity package [16], which provides a native code implementation for the AprilTag tracker. This integration significantly enhances the accuracy of positioning virtual elements within our AR application, ensuring a seamless and realistic interaction for users.

For the AR interactions, we employed Unity's Visual Effect Graph (VFX Graph), a powerful tool within Unity that allows developers to create highly customizable particle effects and visual effects. This tool was instrumental in creating the virtual sphere, offering users an innovative and immersive way to engage with and comprehend weather information.

### 3.3 Localization

As shown in Figure 3, to ensure the precise overlay of the model onto the physical replica of Taiwan Island, the localization framework is divided into two core modules. The first module, represented in blue within the system architecture, leverages the AprilTag Detection system, employing the TagStandard41h12 specification (Figure 4). This advanced detection method scans for specific tag IDs, allowing it to discern the precise location and orientation of the tags placed on the physical model.

The second module, delineated in red, is built upon the robust capabilities of AR Foundation, which serves as a common interface for augmented reality development across iOS and Android platforms. It incorporates the plane detection feature, enabling the application to understand and map the topology of surfaces in the real world. This seamless integration ensures that the virtual model is accurately displayed on the user's device, enhancing the realism and interactivity of the AR experience.

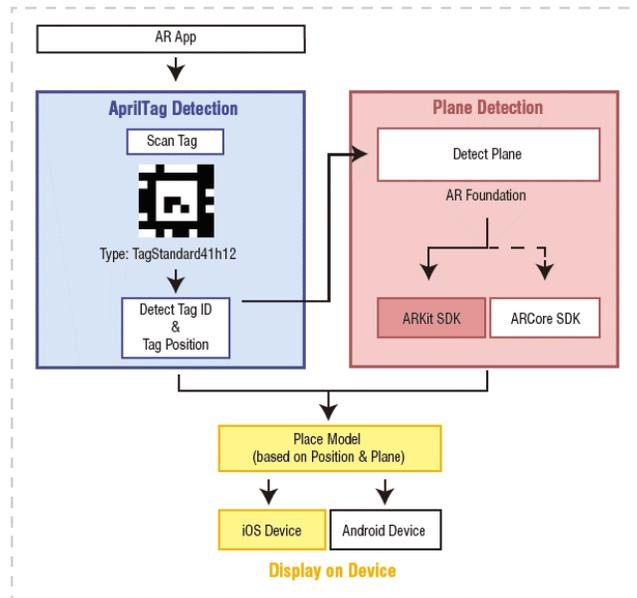

Figure 4: The Flow of Localization

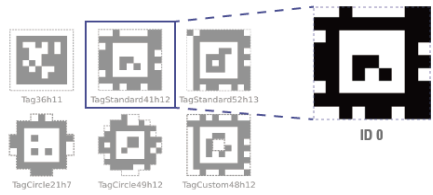

Figure 3: The Types of AprilTag

### 3.3.1 AprilTag Detection

AprilTag Detection is a visual fiducial system commonly used for a variety of applications including robotics and augmented reality for precise localization. The system uses square barcodes, known as AprilTags, which are easily detectable and can be uniquely identified. The AprilTag frame are illustrated as Figuare 5, when these tags are detected by a camera or sensor, the system recognizes the tag ID and calculates the relative position and orientation of the tag in space. This allows for the accurate placement of digital content in the physical world when using augmented reality, as it provides a fixed reference point in the environment.

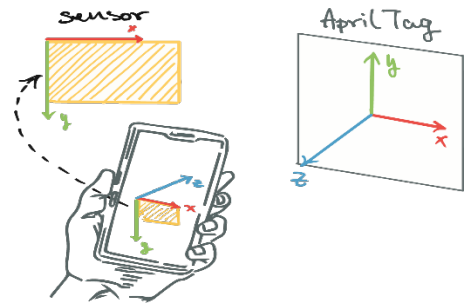

Figure 5: AprilTag Frame Transfoems

### 3.3.2 Plane Detection

Plane Detection is a feature of augmented reality platforms that allows the detection of horizontal and vertical surfaces in the real world. In the context of AR development, it is supported by AR Foundation, a cross-platform framework used for building AR experiences. It supports both ARKit for iOS devices and ARCore for Android devices. As shown in Figure 6, by detecting planes, the AR application can understand the layout of the environment and place digital objects on flat surfaces like tables, floors, or walls.

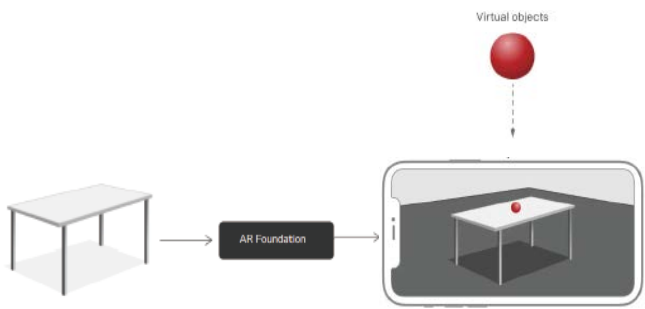

Figure 6: Diagram of Plane Detection

### 3.4 Data Visualization

Our data visualization system operates through two main frameworks. The first, marked by the blue area, is tasked with gathering real-time data from backend APIs, which it receives from sources such as the Central Weather Bureau (CWB) and the Environmental Protection Agency (EPA). It processes and serializes this information into JSON format, which is selected for its efficiency and the ease it provides in the transmission of current and readily usable data.

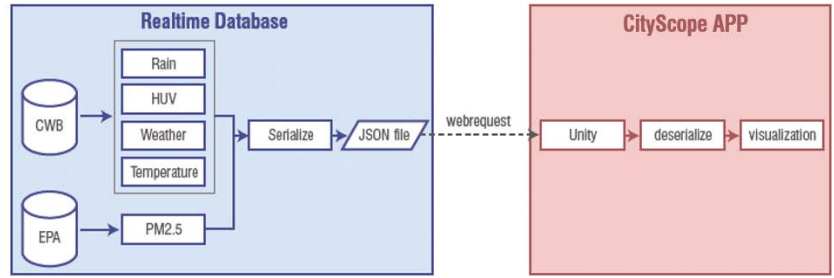

Figure 7: The Flow of Data Visualization

The second framework, indicated by the red area, focuses on data transmission and visualization. It uses web requests to send the JSON data wirelessly to our CityScope APP running the Unity engine. In Unity, the data undergoes a process called JSON deserialization, converting it into a format suitable for visual representation. This step is crucial for turning complex data into understandable and interactive visual elements, enhancing the user's ability to engage with and comprehend the information.

*3.4.1 HUV*

The visualization of ultraviolet (UV) effects are designed based on the UV index, with particle colors also planned according to the color scheme of the UV index. The index ranges from 0 to 11+, with colors transitioning from green to purple to indicate the level of UV exposure, from low to high.

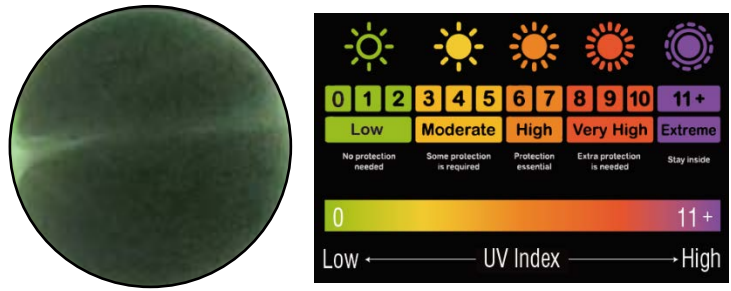

Figure 8: Low UV Visualization

*3.4.2 Temperature*

The visualization of temperature effects aims to represent the intensity of convection currents. The color scheme ranges from blue to indicate cold temperatures to red for hot temperatures. The stronger the convection, the colder the temperature is represented, and conversely, weaker convection corresponds to hotter temperatures.

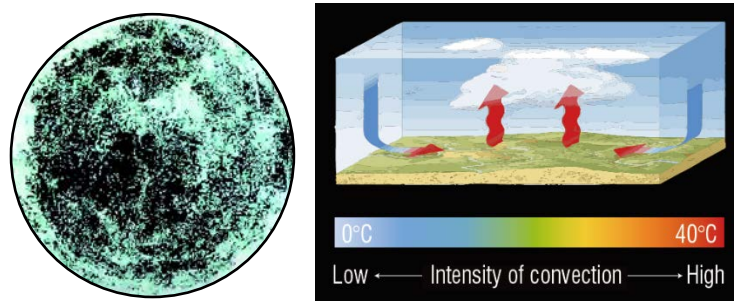

Figure 9: Low Temperature Visualization

*3.4.3 PM2.5*

The visualization of PM2.5 effects are designed to to represent the levels of fine particulate matter concentration in the air, using Air Quality Index (AQI) values as a guideline. A higher concentration is depicted through a darker brown shade, whereas a lower concentration is shown using a lighter green shade.

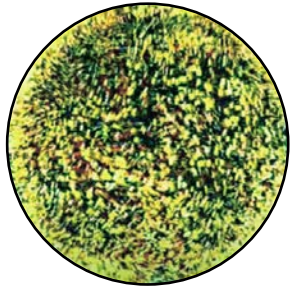

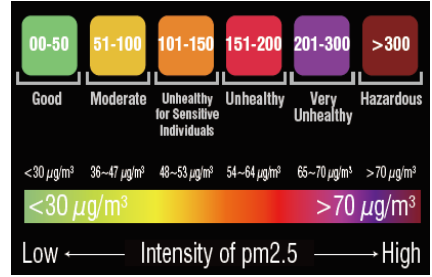

Figure 10: Low PM2.5 Visualization

*3.4.4 Rainfall*

The visualization of rain effects is influenced by the real-time rainfall data received. The higher the numerical value of the rainfall, the denser the particle effects for rain appear, and conversely, lower values result in sparser rain effects, mimicking light drizzles or occasional raindrops.

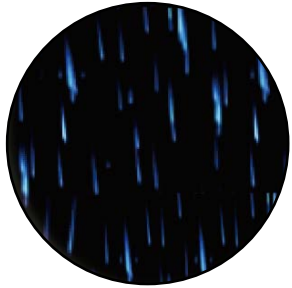

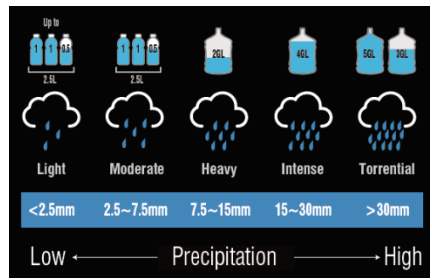

Figure 11: Light Rainfall Visualization

## 4 EXPERIMENTAL RESULTS

### 4.1 Positional Analysis

Figure 12 illustrates the conventional AprilTag Detection, which requires continuous alignment between the camera and the tag to ensure that virtual objects remain stable and do not shift or disappear. This is in stark contrast to traditional Plane Detection methods, where users must manually place objects on surfaces, often leading to less precise positioning of virtual elements.

However, our experiment innovatively combines AprilTag Detection with Plane Detection to address these challenges. In this study, we have developed a system that merges the advantages of Marker-Based AR positioning with the benefits of Markerless AR, which maintains the fixed point even when the camera moves. This combination enhances both accuracy and stability. We pass the AprilTag-derived position data to plane detection, enabling the precise placement of virtual models and city pins on the physical Taiwan replica. This is illustrated in Figure 12, where the virtual Taiwan representation is hidden, leaving only the city pins visible. This integration significantly improves the accuracy of AR object placement, making our system more reliable and user-friendly.

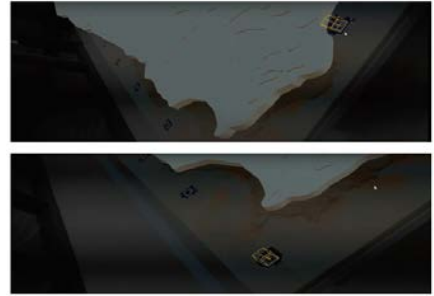

Figure 12: Showcase of AprilTag Detection

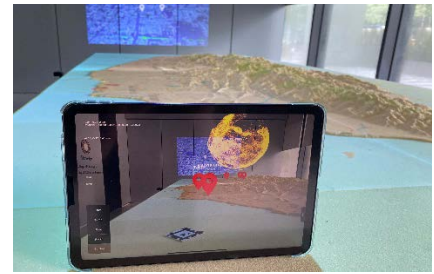

Figure 13: Showcase of Our Localization Approach

### 4.2 User Interface

In the user interface aspect, as shown in Figure 13, this study has designed four buttons for four meteorological pieces of information. Users begin by clicking on the city pin objects they wish to preview. Subsequently, they can click on the desired meteorological information to gain insights. Relevant numerical values will be visualized within a virtual sphere above the Taiwan Island replica.

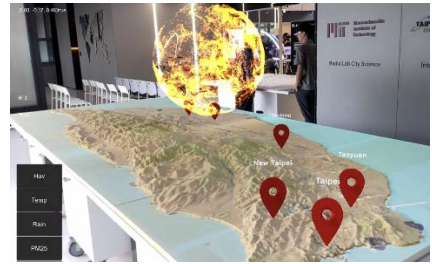

Figure 14: User Interface in AR APP

### 5 CONCLUSION

In this study, we have successfully demonstrated how Augmented Reality (AR) can transform the visualization and interpretation of complex meteorological data. Our integration of advanced AR localization with an innovative data visualization system provides an interactive and intuitive interface for both the public and urban decision-makers. The use of AprilTag Detection in conjunction with Plane Detection, as evidenced in our experimental results, markedly improves the accuracy of virtual object placement, effectively overcoming the limitations of traditional AR methods. This innovative approach enhances the user's ability to comprehend and interact with intricate weather data in a contextually relevant and spatially accurate manner.

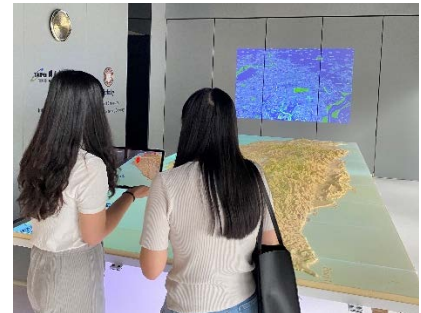

Figure 15: Public participation in using CityScope AR APP

Our research not only advances the field of data visualization in meteorology but also significantly enhances public engagement with complex information. The user-friendly interface of our AR tool empowers citizens to interact with and understand weather data more intuitively. By making meteorological data accessible and understandable to a wider audience, our study paves the way for more inclusive and participatory approaches in managing and responding to environmental challenges. The potential of AR to democratize access to complex information holds great promise for future developments across various domains. The developed AR tool not only facilitates a deeper understanding of meteorological information but also aids in informed decision-making in urban planning and environmental management.

### ACKNOWLEDGMENTS

I would like to extend my sincere thanks to Michael Lin, Guest Director of City Science Lab @ Taipei, and the entire team at City Science Lab, for their invaluable guidance and support throughout this project.